\documentstyle[preprint,aps,epsfig]{revtex}

\def\be{\begin{equation}}
\def\ee{\end{equation}}
\def\ep{\epsilon}
\def\t{\tilde}
\def\sg{\sigma}
\def\et{\tilde{\epsilon}}

\begin{document}
\draft
\title{Electrorotation of a pair of spherical particles}
\author{J. P. Huang$^1$, K. W. Yu$^1$ and G. Q. Gu$^{1,2}$}
\address{$^1$Department of Physics, The Chinese University of Hong Kong,
 Shatin, NT, Hong Kong}
\address{$^2$Information College, East China Normal University,
 Shanghai 200 062, China}
\maketitle

\begin{abstract}
We present a theoretical study of electrorotation (ER) of two spherical 
particles under the action of a rotating electric field. 
When the two particles approach and finally touch, the mutual polarization
interaction between the particles leads to a change in the dipole moment 
of the individual particle and hence the ER spectrum, as compared to that 
of the well-separated particles. The mutual polarization effects are 
captured by the method of multiple images.
From the theoretical analysis, we find that the mutual polarization effects
can change the characteristic frequency at which the maximum angular 
velocity of electrorotation occurs. The numerical results can be 
understood in the spectral representation theory.

\end{abstract}
\vskip 5mm

\section{Introduction}

When a suspension of colloidal particles or biological cells is exposed 
to an external electric field, the analysis of the frequency-dependent 
response yields valuable information on various processes, like the 
structural (Maxwell-Wagner) polarization effects \cite{Gimsa,Gimsa99}.
For most substances, the permittivity and conductivity are only constant
over a limited frequency range. The general tendency is for permittivity
to decrease, and conductivity to concomitantly increase, in a series of 
steps as frequency increases. These step changes are called dispersions.

While the frequency-dependent response of biological cells can be 
investigated by the method of dielectric spectroscopy \cite{Asami80}, 
conventional dielectrophoresis and electrorotation (ER) analyze the 
frequency dependence of translations and rotations of single particles in an 
inhomogeneous and rotating external field, respectively \cite{Fuhr,Gimsa91}.
With the recent advent of experimental techniques such as automated video 
analysis \cite{Gasperis} as well as light scattering methods \cite{Gimsa99}, 
the motion of particles can be accurately monitored.
In ER, the rotating field induces a cell dipole moment which rotates at the
angular frequency of the external field. Any dispersion process causes a
spatial phase shift between the induced dipole moment and the external field
vector, giving rising to a torque which causes the rotation of individual 
cells \cite{Burt}.

In the dilute limit in which a small volume fraction of particles or cells
are suspended in a medium while the particles are well separated, 
we can concentrate on the ER response of individual particles by ignoring 
the mutual interactions between the particles.
However, if the suspension is non-dilute, one may not ignore the 
interactions.
Moreover, when the strength of the rotating electric field increases, 
the Brownian motion can be ignored and the polarized particles tend to 
aggregate along the rotating field even in the dilute limit.

As an initial model, we will consider a pair of interacting particles 
dispersed in a suspension.
When the two particles approach and finally touch, the mutual polarization 
interaction between the particles leads to a change in the dipole moment
of individual particles and hence the ER spectrum, as compared to that 
of the well-separated particles. We will present a theoretical study on the 
ER spectrum of two spherical particles in the presence of a rotating 
electric field, and capture the mutual polarization effects by the method 
of multiple images \cite{Yu}, recently developed to calculate the 
interparticle force between touching spherical particles.

\section{Multiple image method for a pair of dielectric spheres}

We first consider an isolated spherical cell of complex permittivity 
$\et_1=\ep_1+\sg_1/(i\omega)$ dispersed in a suspension medium of 
$\et_2=\ep_2+\sg_2/(i\omega)$, where $\omega$ is the frequency of the 
external field $\vec{E}_0$, and $i=\sqrt{-1}$.
If one applies a rotating electric field, there is in general a phase 
difference between the induced dipole moment and the applied field vector, 
and thus a net torque is acting on the cell. 
The spin friction (analogous to the Stokes drag for translational motions) 
causes a steady-state rotational motion. 
In this case, the dipole moment of the isolated cell $\t{p}$ is
\be
\t{p}=\frac{1}{8}\et_2 D^3 bE_0,
\ee 
where $b=(\et_1-\et_2)/(\et_1+2\et_2)$ is the dipole factor, 
and $D$ the diameter of the cell. 

To account for the effect of multipole interaction on the electrorotation, 
we consider a pair of touching spherical cells at a separation $R$ 
suspended in a medium.
We will calculate the effect of multipole interaction on the dipole moment.
For a rotating field, the total dipole moment of one cell of the pair 
is given by 
\be
\t{p}^*=\t{P}_T\langle \cos ^2\theta\rangle+\t{P}_L
  \langle \sin ^2\theta\rangle
 =\frac{1}{2}(\t{P}_T+\t{P}_L), 
\ee 
where $\theta$ is the angle between the dipole moment and the line 
joining the centers of the spheres. The average over all possible 
orientations of the cells gives a factor 1/2. 
In Eq.(2), $\t{P}_T$ ($\t{P}_L$) is the transverse (longitudinal) 
dipole moment, being perpendicular (parallel) to the line joining 
the centers of the spheres. 
The transverse and longitudinal dipole moments are given by \cite{Yu}
\begin{eqnarray}
\t{P}_L&=&\t{p}\sum_{n=0}^\infty(2b)^n \left(\frac{\sinh \alpha}
 {\sinh (n+1)\alpha}\right)^3,\nonumber\\
\t{P}_T&=&\t{p}\sum_{n=0}^\infty(-b)^n \left(\frac{\sinh \alpha}
 {\sinh (n+1)\alpha}\right)^3,\nonumber 
\end{eqnarray}
where $\alpha$ satisfies the relation $\cosh \alpha=R/D$.
For two touching particles, the dipole moment of a particle is given by 
\be
\t{p}^*=\frac{1}{8}\et_2 D^3b^*E_0,
\ee 
where the dipole factor of a pair is given by:
\be
b^*=\frac{8\t{p}^*}{\et_2 D^3E_0}
=\frac{1}{2}b \left[\sum_{n=0}^\infty(2b)^n \left(\frac{\sinh \alpha}
 {\sinh (n+1)\alpha}\right)^3+
 \sum_{n=0}^\infty(-b)^n 
 \left(\frac{\sinh \alpha}{\sinh (n+1)\alpha}\right)^3\right].
\ee 

It is known that the angular velocity of electrorotation of a particle
is related to the dipole factor ($b$ or $b^*$) as follows
\be
\Omega=-F(\ep_2,\eta,E_0)Im[Dipole\ Factor]
\ee 
where $F$ is a function of $\ep_2$, the dynamic viscosity $\eta$ of the 
medium, as well as the electric-field magnitude $E_0$, and $Im[\cdots]$ 
denotes the imaginary part of $[\cdots]$.
For a spherical particle, $F(\ep_2,\eta,E_0)=\ep_2 E_0^2/2\eta$.
Regarding $\eta$, the spin friction expression suffices for an isolated 
particle. However, for two interacting particles, we must consider the 
more complicated suspension hydrodynamics.

\section{Spectral representation and dispersion spectrum}

In a recent paper \cite{Lei}, we studied the dielectric behavior of cell 
suspensions by employing the Bergman-Milton spectral representation of 
the effective dielectric constant \cite{Bergman}. 
By means of the spectral representation, we derived the dielectric 
dispersion spectrum in terms of the electrical and structure parameters 
of the cell models. 
The spectral representation is a rigorous mathematical formalism of 
the effective dielectric constant of a composite material.
The essence of the spectral representation is to define the following 
transformations. If we denote a complex material parameter
 \be
\t{s}=\left(1 - {\t{\ep}_1\over \t{\ep}_2} \right)^{-1},
 \ee
then the dipole factor $b^*$ admits the general form
 \be
b^* = \sum_n {F_n \over \t{s}-s_n},
 \ee
where $n$ is a positive integer, i.e., $n=1, 2, ...$, and $F_n$ and $s_n$ 
are the $n$-th microstructure parameters of the composite material 
\cite{Bergman}.

Thus the spectral representation offers the advantage of 
the separation of material parameters 
(namely the dielectric constant and conductivity) from the cell
structure information, thus simplifying the study. 
From the spectral representation, one can readily derive the dielectric 
dispersion spectrum, with the dispersion strength as well as the 
characteristic frequency being explicitly expressed in terms of the 
structure parameters and the materials parameters of the cell suspension
\cite{Lei}.
The actual shape of the real and imaginary parts of the permittivity
over the relaxation region can be uniquely determined by the Debye 
relaxation spectrum, parametrized by the characteristic frequencies 
and the dispersion strengths. So, we can study the impact of these 
parameters on the dispersion spectrum directly.
In what follows, we further express the dipole factor $b$ and $b^*$ 
in the spectral representation. 
The dipole factor $b$ admits the form $b=F_1/(\t{s}-s_1)$
in the spectral representation, where $F_1=-1/3$ and $s_1=1/3$. 
For the dielectric dispersion spectrum, we further define two dimensionless 
real parameters \cite{Lei} 
$$
s=(1-\ep_1/\ep_2)^{-1},\ {\rm and} \ \  t=(1-\sg_1/\sg_2)^{-1} ,
$$
where $s$ and $t$ are the dielectric and conductivity contrasts respectively.
After simple manipulations, Eq.(7) becomes
\be
b=\frac{F_1}{s-s_1}+\frac{\delta \ep_1}{1+if/f_{c}}
\ee 
where $f_c$ is the characteristic frequency, at which the maximum angular
velocity of electrorotation occurs, and $\delta\ep_1$ is the dispersion 
magnitude \cite{Lei}: 
\begin{eqnarray}
\delta\ep_1&=&F_1\frac{s-t}{(t-s_1)(s-s_1)},\nonumber\\
f_{c}&=&\frac{1}{2\pi}\frac{\sg_2}{\ep_2}\frac{s(t-s_1)}{t(s-s_1)}.
\end{eqnarray}

Similarly, the dipole factor of a pair of interacting particles $b^*$ 
can be rewritten in the spectral representation as
\be
b^*=\sum_{m=1}^{\infty}\left( \frac{F_{m}^{(T)}}{\t{s}-s_{m}^{(T)}}
 +\frac{F_{m}^{(L)}}{\t{s}-s_{m}^{(L)}} \right)
\ee 
where
\begin{eqnarray}
F_{m}^{(T)}&=&F_{m}^{(L)}=-\frac{2}{3}m(m+1)\sinh ^3\alpha 
 e^{-(2m+1)\alpha},\nonumber\\
s_{m}^{(T)}&=&\frac{1}{3}[1+e^{-(1+2m)\alpha}], \ \ 
s_{m}^{(L)}=\frac{1}{3}[1-2e^{-(1+2m)\alpha}] . \nonumber
\end{eqnarray}
In the above derivation, we have used the identity
$$
{1\over \sinh^{3} x} = \sum_{m=1}^{\infty}4m(m+1)e^{-(1+2m)x}.
$$
We should remark that the spectral representation of $b^*$ [Eq.(10)] is an 
exact transformation \cite{Thorpe} of the multiple image expression [Eq.(4)].
Thus, the spectral representation of $b^*$ consists of a discrete set of 
simple poles (see Fig.5 below). 
The $m=1$ pole $s_{1}^{(L)}$ ($s_{1}^{(T)}$) for the longitudinal 
(transverse) field case deviates significantly from $s=1/3$ while 
$s_{1}^{(L)}$ becomes small in the touching limit, $\alpha \to 0$. 
In what follows, we will show that this pole gives a significant 
contribution to the ER spectrum at low frequency.
As $m$ increases, however, the series $s_{m}^{(T)}$ and $s_{m}^{(L)}$ 
accummulate to $s=1/3$, giving rise to a dominant contribution near the 
isolated-sphere frequency $f_c$. Moreover, each term in the spectral 
representation expression of $b^*$ can be rewritten as, 
\begin{eqnarray}
\frac{F_{m}^{(T)}}{\t{s}-s_{m}^{(T)}}&=&\frac{F_{m}^{(T)}}{s-s_{m}^{(T)}}
 +\frac{\delta\ep_{m}^{(T)}}{1+if/f_{mc}^{(T)}},\nonumber\\
\frac{F_{m}^{(L)}}{\t{s}-s_{m}^{(L)}}&=&\frac{F_{m}^{(L)}}{s-s_{m}^{(L)}}
 +\frac{\delta\ep_{m}^{(L)}}{1+if/f_{mc}^{(L)}},\nonumber
\end{eqnarray}
where $\delta\ep_{m}^{(T)}$ (or $\delta\ep_{m}^{(L)}$) and $f_{mc}^{(T)}$ 
(or $f_{mc}^{(L)}$) are the dispersion magnitudes and the characteristic 
frequencies of the transverse (longitudinal) field cases, 
obtained respectively by replacing $F_1$ and $s_1$ in the expressions of 
$\delta \ep_1$ and $f_{c}$ in Eq.(9) with 
$F_{m}^{(T)}$ (or $F_{m}^{(L)}$) and $s_{m}^{(T)}$ (or $s_{m}^{(L)}$).
Note that $f_{1c}^{(L)}$ is significantly lower than $f_c$. 
The ER spectrum of two spheres will consist of a series of sub-dispersions.
In what follows, we will investigate the effect of multipole interaction 
on the electrorotation of cell, as shown in the next section.

\section{Numerical results}

For convenience, we let $F(\ep_2,\eta,E_0)=1$ in our numerical calculation, 
both for the isolated-particle and touching-particle cases, 
in order to study the effect of multipole interaction on the ER spectrum.
It is because we will focus on the frequency dependence of the ER spectrum.
On the other hand, we do not know the dynamic viscosity for two approaching
spheres.

Fig.1 is plotted to compare the two cases for three different conductivity 
contrast $t$.
It is evident that, for both cases, a small $|t|$ yields a high 
characteristic frequency, at which the peak locates. Moreover, a smaller 
$|t|$ gives a larger peak value. Given a constant $t$, the touching-cell 
case predicts a shift of the peak location to lower frequency than that 
predicted by the isolated-cell case. At the same time, the peak for the
touching-cell case broaden.
This behavior arises from the effect of the multipole interaction. 
In addition, if $t$ is given, the two cases also predict 
different behavior within the low-frequency region, but almost the same 
behavior within the high-frequency region.
Especially for small $|t|$ (e.g., $t=-1/900$), a second peak occurs at 
lower frequency. This is in accord with the prediction of the spectral 
representation.

Fig.2 is plotted to exhibit the effect of the material parameter $s$ on 
the two cases. It is observed that, for both cases, a larger $s$ gives 
a larger peak value. 
Moreover, the peak of different $s$ appears at almost the same frequency 
for both cases. In summary, $s$ may affect the peak value, but not the 
peak location.

Fig.3 is plotted for three different values of the medium conductivity 
$\sg_2$ at $s=1.1$ and $t=-1/90$. 
For a larger $\sg_2$, we get a larger characteristic frequency at which 
the peak occurs. 
It is evident that $\sg_2$ may affect the peak location, but not the peak 
value.

In Fig.4, as $R/D\approx 1$ (e.g., $R/D=1.0333$), the deviation between 
the two cases is evident. In other words, the multipole interaction does 
play an important role in the ER spectrum and the effect cannot be 
neglected for touching particles. However, as the separation increases, 
say, $R/D=2$, both cases predicts the same ER spectrum. From the results,
we would say that the effect of multipole interaction may be neglected 
at $R/D > 2$.

In Fig.5, we plot the spectral representation for the two sets of poles 
$s_{m}^{(L)}$ (or $s_{m}^{(T)}$) for $m=1$ to $100$.
As $m$ increases, $s_{m}^{(L)}$ increases up to $1/3$, while $s_{m}^{(T)}$ 
decreases towards $1/3$.
For small $R/D$ ratio (say, $R/D=1.0333$), 
the longitudinal field plays an more important role in determining 
electrorotation spectrum than the transverse field. 
However, if $R/D$ ratio is large enough, say, $R/D \ge 2$, 
(results not shown here), the effect of multipole interaction is negligible
both for the longitudinal and transverse field cases and the results 
for an isolated cell recover.
In summary, as the two cells approach and finally touch, one must take into 
account the effect of multipole interaction on the electrorotation of cells.

\section*{Discussion and Conclusion}

Here a few comments on our results are in order.
In this work, we attempted a theoretical study of the electrorotation of 
two approaching spherical particles in the presence of a rotating electric 
field. 
When the two particles approach and finally touch, the mutual polarization
interaction between the particles leads to a change in the dipole moment 
of individual particles and hence the ER spectrum, as compared to that 
of the well-separated particles. The mutual polarization effects are studied
via the multiple image method. From the results, we find that the mutual 
polarization effects can change the characteristic frequency substantially.
We should remarked that the multiple image method for two dielectric 
spheres is only approximate \cite{Choy}, but the approximation is quite 
good \cite{Yu}. 
More accurate calculations based on bispherical coordinates can be attempted. 
However, we believe that similar conclusions can be obtained.

Moreover, when the volume fraction of the suspension is large, the 
rotating electric field leads to plate-like aggregation of particles in 
the plane of the applied field. In this case, the electrorotation 
spectrum can be modified further due to the local field effects arising 
from the many-particle system. To this end, a first-principles approach
can be used to handle the many-particle and multipole interactions
\cite{Yu2000}.

Regarding the dynamic viscosity, the spin friction expression for 
an isolated spherical cell suffices. 
However, for many cells interacting in a suspension, we must consider the 
hydrodynamics. Regarding the dynamic viscosity for two touching spheres, 
we must consider the suspension hydrodynamics of two approaching spheres.
This is a formidable task and should be a topic for future work.

\section*{Acknowledgments}
This work was supported by the Research Grants Council of the Hong Kong 
SAR Government under grant CUHK 4245/01P. 
G.Q.G. acknowledges the financial support from a Key Project of the 
National Natural Science Foundation of China under grant 19834070.
K. W. Y. acknowledges useful correspondence with Professor Mike Thorpe 
regarding the spectral representation of two approaching spheres, and 
extensive discussion with Professor Tuck Choy on the multiple image method.

\newpage

\begin{figure}[h]
\caption{The minus dipole factor (being proportional to the angular 
velocity of electrorotation) plotted versus frequency both for an 
isolated cell (dashed lines) and touching cells (solid lines) 
for three different conductivity contrasts $t$ 
at $s=1.1$ and $\sg_2=2.8 \times 10^{-4}Sm^{-1}$.}
\end{figure}

\begin{figure}[h]
\caption{Similar to Fig.1 but for three different dielectric contrasts $s$ 
at $t=-1/90$ and $\sg_2=2.8 \times 10^{-4}Sm^{-1}$ for an isolated cell 
(lines) and touching cells (symbols).}
\end{figure}

\begin{figure}[h]
\caption{Similar to Fig.1 but for three different medium conductivities 
$\sg_2$ at $s=1.1$ and $t=-1/90$.}
\end{figure}

\begin{figure}[h]
\caption{Similar to Fig.1 but for three different separation ratios $R/D$ 
at $s=1.1$, $t=-1/900$ and $\sg_2=2.8 \times 10^{-4}Sm^{-1}$.}
\end{figure}

\begin{figure}[h]
\caption{The pole spectrum for two approaching spheres for several 
separation ratios $R/D$ for the longitudinal (open symbols) and
transverse (filled symbols) cases. Note that the longitudinal (transverse) 
part of the spectrum are on the small (large) $s$ side of the spectrum.
The lines are guides to the eyes.}
\end{figure}

\newpage
\centerline{\epsfig{file=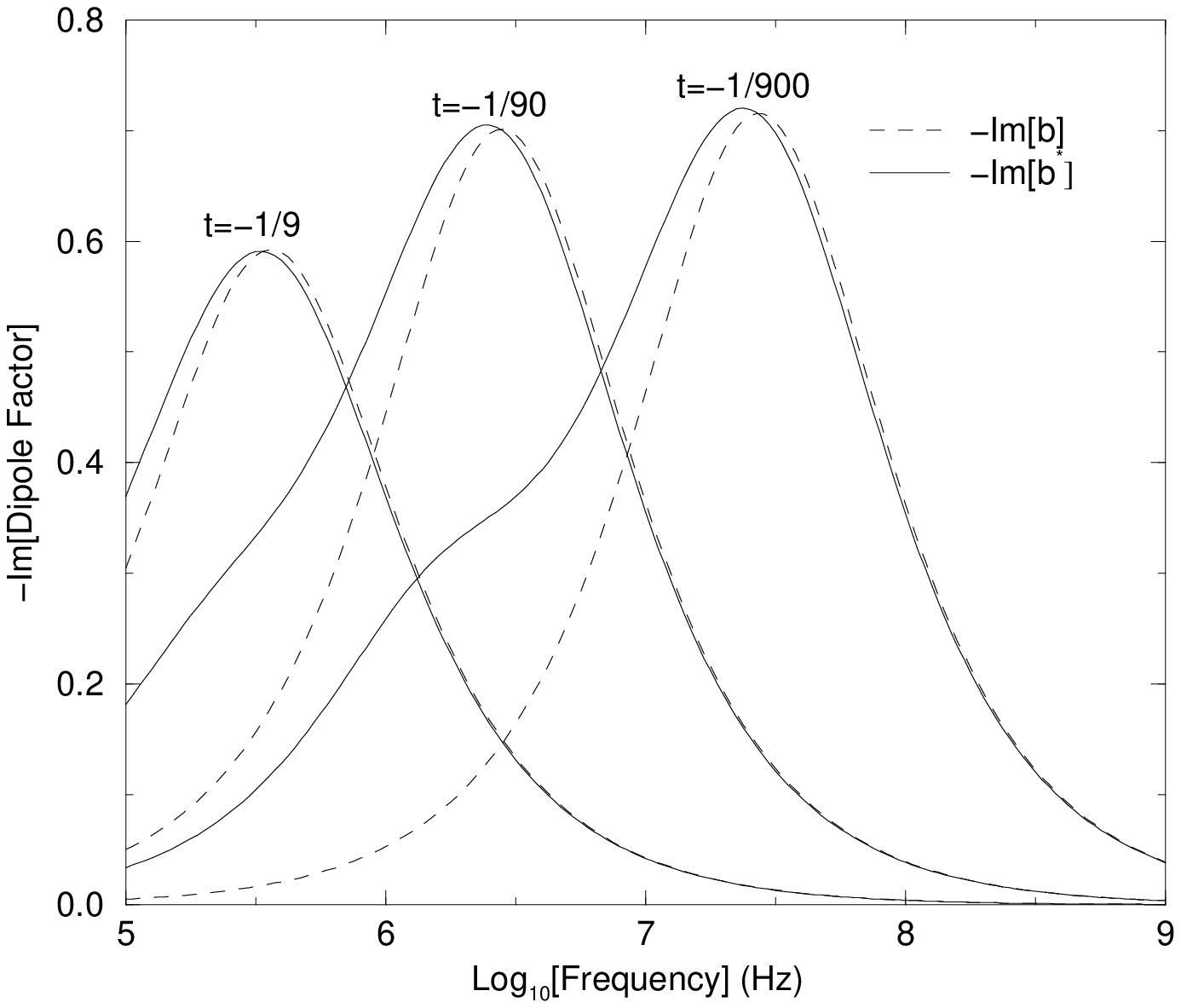,width=\linewidth}}
\centerline{Fig.1}

\centerline{\epsfig{file=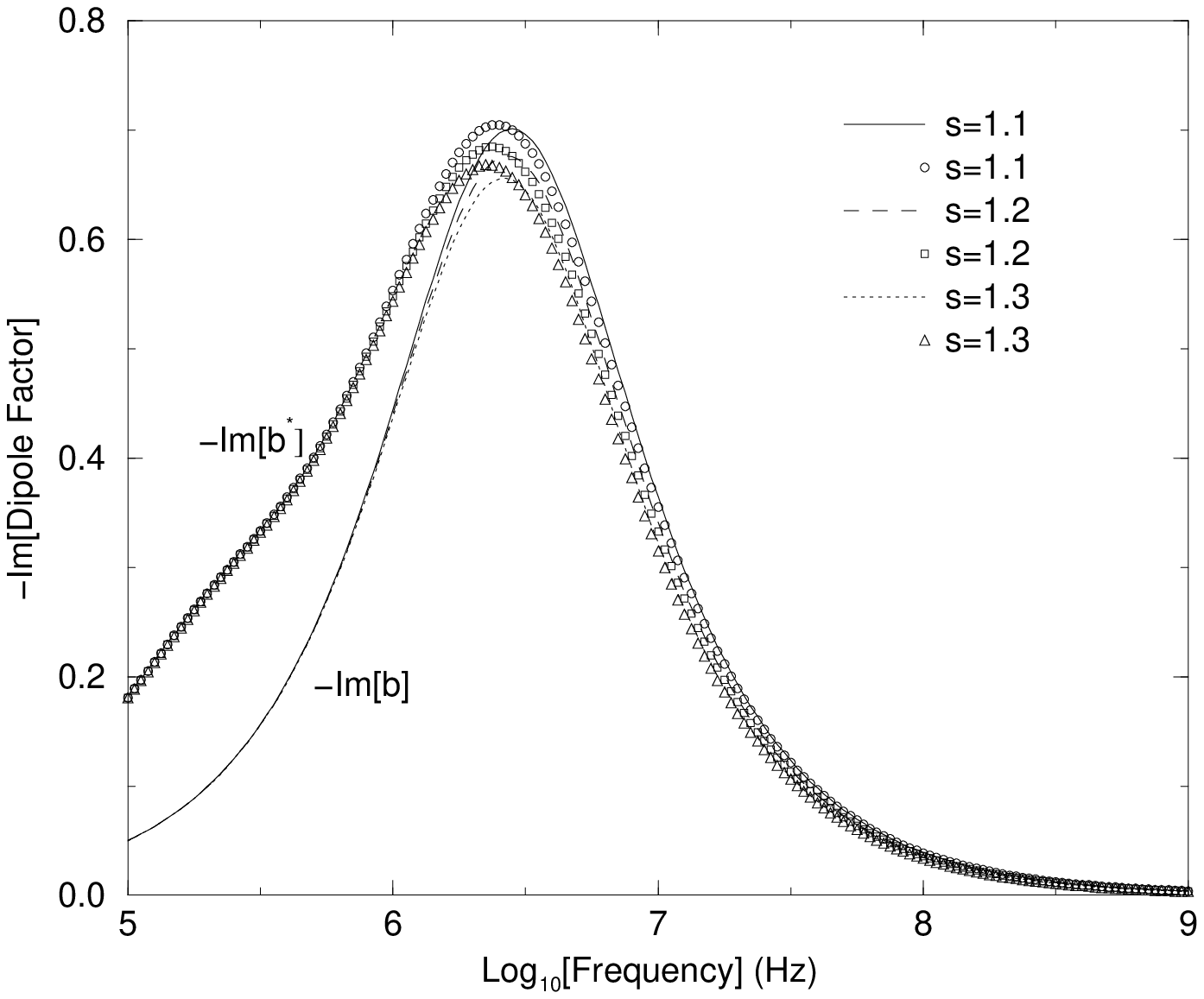,width=\linewidth}}
\centerline{Fig.2}

\centerline{\epsfig{file=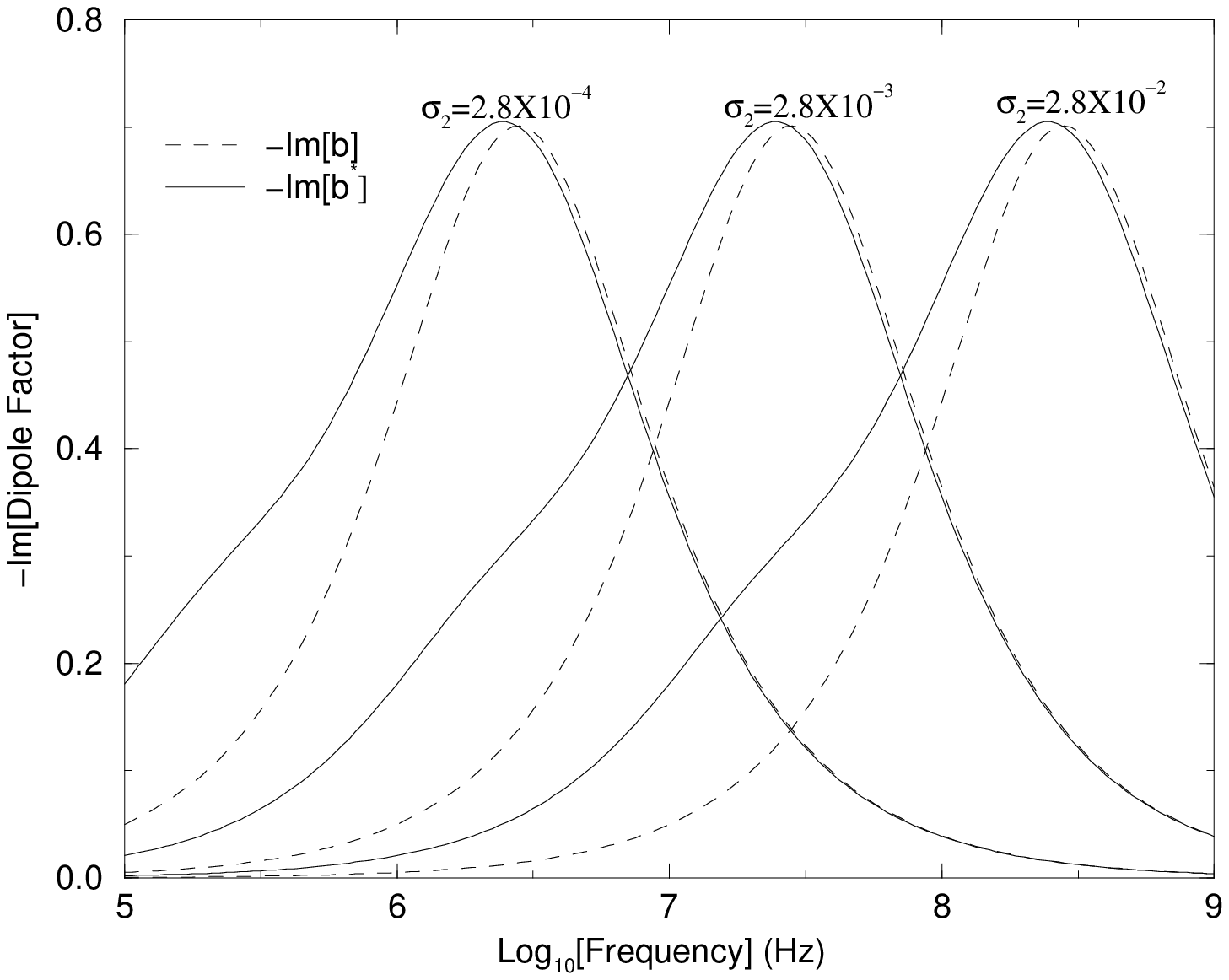,width=\linewidth}}
\centerline{Fig.3}

\centerline{\epsfig{file=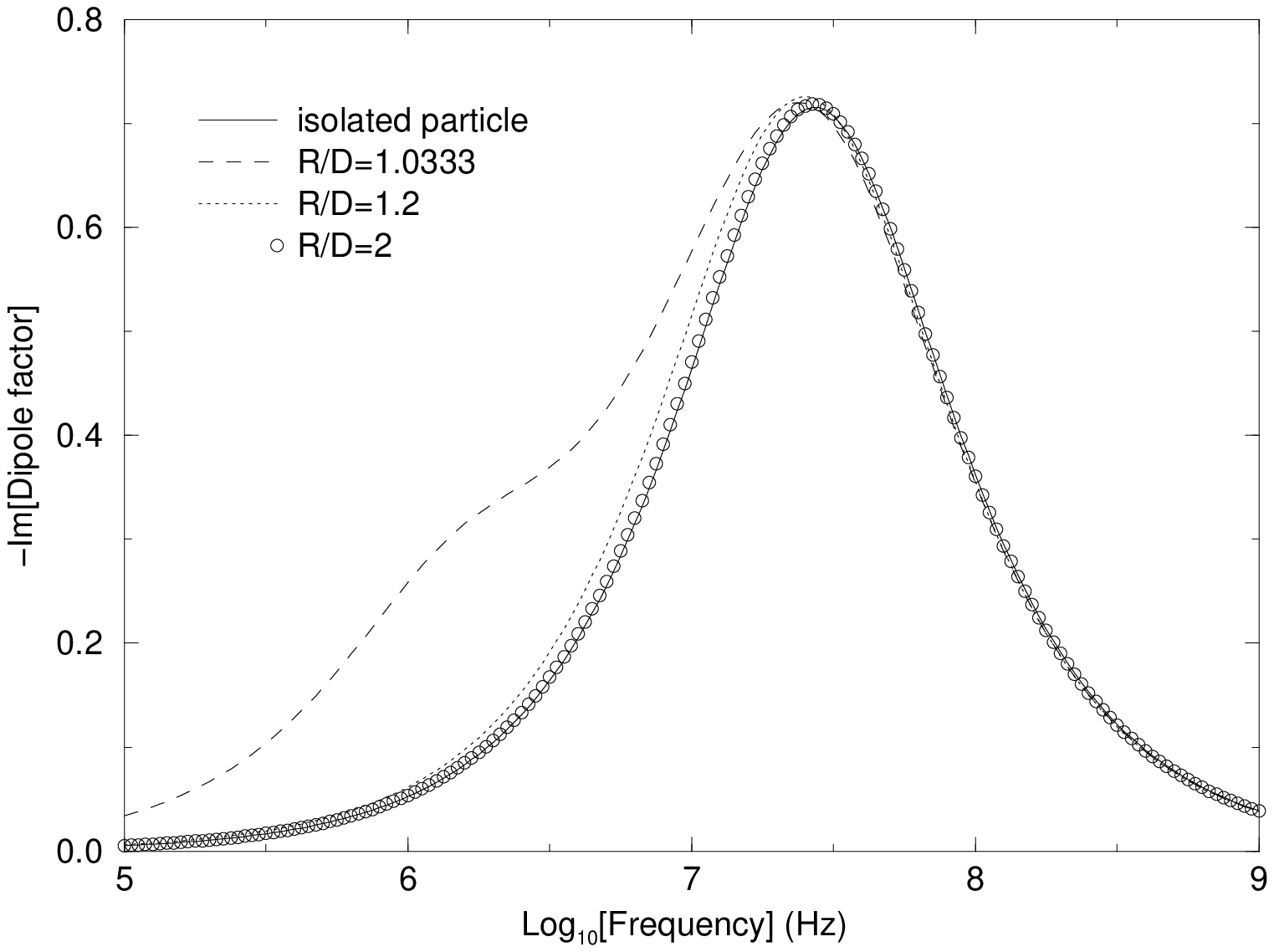,width=\linewidth}}
\centerline{Fig.4}

\centerline{\epsfig{file=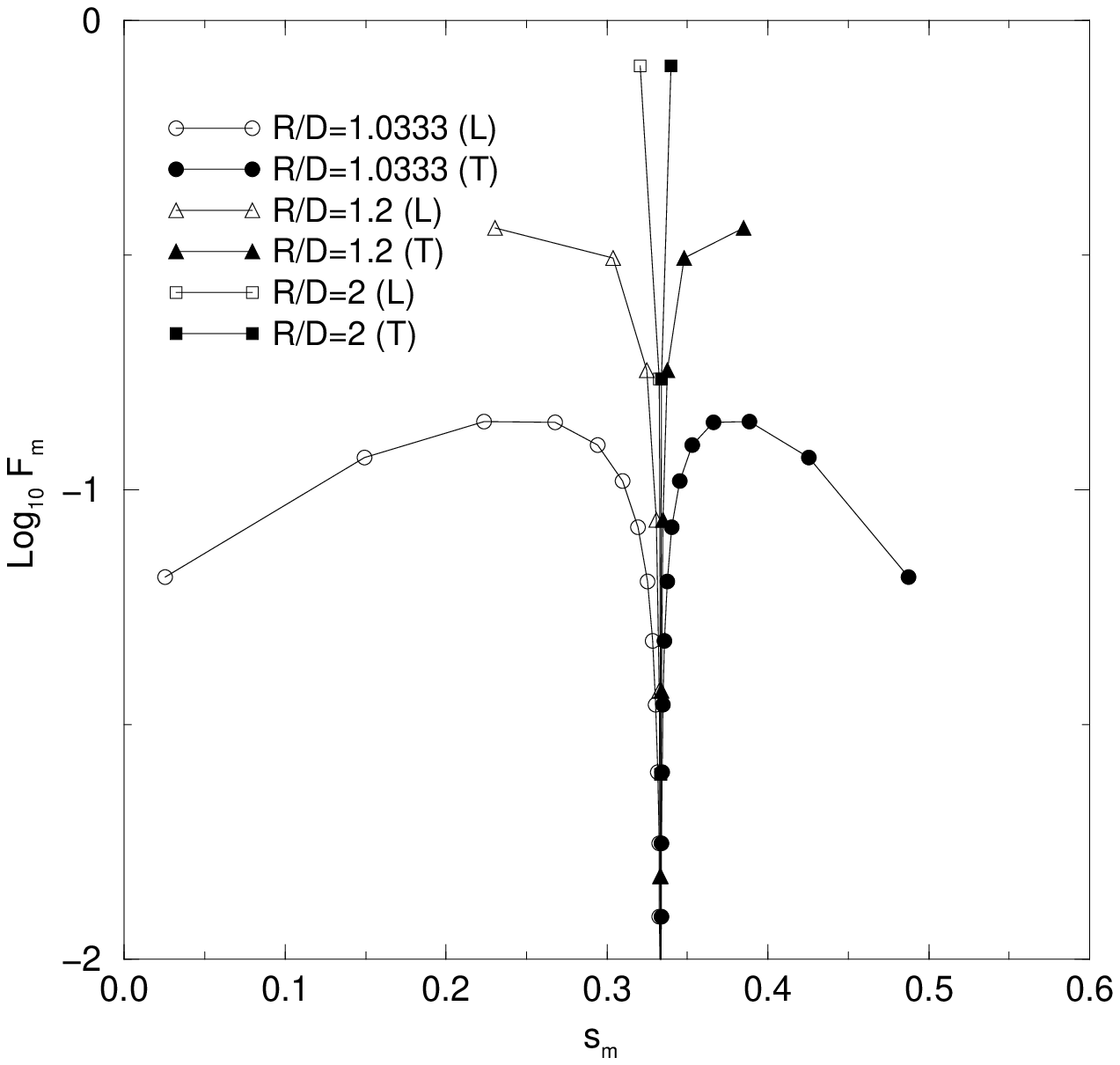,width=\linewidth}}
\centerline{Fig.5}


\begin{references}

\bibitem{Gimsa} For a review, see J. Gimsa and D. Wachner,
 Biophys. J. {\bf 77}, 1316 (1999).
\bibitem{Gimsa99} J. Gimsa, Ann. NY Acad. Sci. {\bf 873}, 287 (1999).
\bibitem{Asami80} K. Asami, T. Hanai and N. Koizumi, Jpn. J. Appl. Phys.
 {\bf 19}, 359 (1980).
\bibitem{Fuhr} G. Fuhr, J. Gimsa and R. Glaser, Stud. Biophys. {\bf 108},
 149 (1985).
\bibitem{Gimsa91} J. Gimsa, P. Marszalek, U. Lowe and T. Y. Tsong, 
 Biophys. J. {\bf 73}, 3309 (1991).
\bibitem{Gasperis} G. De Gasperis, X.-B. Wang, J. Yang, F. F. Becker 
 and P. R. C. Gascoyne, Meas. Sci. Technol. {\bf 9}, 518 (1998).
\bibitem{Burt} J. P. H.  Burt, K. L. Chan, D. Dawson, A. Parton and 
 R. Pethig, Ann. Biol. Clin. {\bf 54}, 253 (1996); see also 
 http://www.ibmm.informatics.bangor.ac.uk/pages/science/rot.htm
 for the basic science of electrorotation.

\bibitem{Yu} K. W. Yu and Jones T. K. Wan, Comput. Phys. Commun. 
 {\bf 129}, 177 (2000).
\bibitem{Lei} Jun Lei, Jones T. K. Wan, K. W. Yu and Hong Sun,
 Phys. Rev. E {\bf 64}, 012903 (2001).

\bibitem{Bergman} D. J. Bergman, Phys. Rep. {\bf 43}, 379 (1978).  
\bibitem{Thorpe} M. F. Thorpe, private communications; 
 see also B. R. Djordjevic, J. H. Hetherington and M. F. Thorpe, 
 Phys. Rev. B {\bf 53}, 14 862 (1996) 
 for a similar method in two dimensions.

\bibitem{Choy} T. C. Choy, A. Alexopoulos and M. F. Thorpe, Proc. R. Soc. 
 (Lond.) A {\bf 454}, 1993 (1998).

\bibitem{Yu2000} K. W. Yu, Hong Sun and Jones T. K. Wan, 
 Physica B {\bf 279}, 78 (2000).
\end{references}
\end{document}